\documentclass[twocolumn,showpacs,preprintnumbers,amsmath,amssymb,nofootinbib]{revtex4}

\usepackage{subfigure}
\usepackage{amsmath,amssymb}
\usepackage{graphicx}
\usepackage{rotating}
\usepackage{bm}
\usepackage{color}

\definecolor{blue}{rgb}{0,0,1}

\definecolor{green}{rgb}{0,1,0}

\definecolor{red}{rgb}{1,0,0}

\newcommand{\be}{\begin{eqnarray}}
\newcommand{\ee}{\end{eqnarray}}

\newcommand{\Gk}{\Gamma_k}
\newcommand{\Tc}{T_{\text{cr}}}
\newcommand{\yb}{\bar\psi}
\newcommand{\xsb}{$\chi$SB}
\newcommand{\Nf}{N_{\text{f}}}
\newcommand{\Nc}{N_{\text{c}}}
\newcommand{\pat}{\partial_t}

\begin{document}

\author{Jens Braun}
\author{Holger Gies}
\affiliation{Institut f\"ur Theoretische Physik, Philosophenweg 16 and %
  19, 69120 Heidelberg, Germany} 
\preprint{HD-THEP-05-26}

\title{Running coupling at finite temperature \\and chiral symmetry
restoration in QCD}

\begin{abstract}

We analyze the running gauge coupling at finite temperature for QCD,
using the functional renormalization group. The running of the
coupling is calculated for all scales and temperatures. At finite
temperature, the coupling is governed by a fixed point of the
3-dimensional theory for scales smaller than the corresponding
temperature.  The running coupling can drive the quark sector to
criticality, resulting in chiral symmetry breaking. Our results
provide for a quantitative determination of the phase boundary in the
plane of temperature and number of massless flavors. Using the
experimental value of the coupling at the $\tau$ mass scale as the
only input parameter, we obtain, e.g., for $\Nf=3$ massless flavors a
critical temperature of $\Tc\approx 148\,\mathrm{MeV}$ in good
agreement with lattice simulations.

\end{abstract}


\maketitle

\section{Introduction}

Strongly interacting matter is believed to have fundamentally
different properties at high temperature than at low or zero
temperature \cite{Karsch:2003jg}. Whereas the latter can be described
in terms of ordinary hadronic states, a hadronic picture at increasing
temperature is eventually bound to fail; instead, a description in
terms of quarks and gluons is expected to arise naturally owing to
asymptotic freedom. In the transition region between these asymptotic
descriptions, effective degrees of freedom, such as order parameters
for the chiral or deconfining phase transition, may characterize the
physical properties in simple terms, i.e., with a simple effective
action \cite{Pisarski:1983ms}.

If a simple description at or above the phase transition does not
exist and the system is strongly interacting in all conceivable sets
of variables \cite{Shuryak:2003xe}, a formulation in terms of
microscopic degrees of freedom has the greatest potential to bridge
wide ranges in parameter space from first principles.

In this Letter, we report a nonperturbative study of
finite-temperature QCD parameterized in terms of microscopic degrees
of freedom: gluons and quarks. We use the functional renormalization
group (RG) \cite{Wegner,Wetterich:yh,bonini} and concentrate on two
problems which are accessible in microscopic language: first, we
compute the running of the gauge coupling driven by quantum as well as
thermal fluctuations, generalizing previous zero-temperature studies
\cite{Gies:2002af}. Second, we investigate the induced quark dynamics
including its back-reactions on gluodynamics, in order to monitor the
status of chiral symmetry at finite temperature. With this strategy,
the critical temperature of chiral symmetry restoration can be
computed. This Letter is particularly devoted to a presentation of the
central physical mechanisms which our study reveals for the dynamics
near the phase transition; all technical details can be found in a
separate publication \cite{BraunGies}.

The functional RG yields a flow equation for the effective average
action $\Gamma_k$ \cite{Wetterich:yh},
\begin{equation}
\pat\Gamma_k=\frac{1}{2}\, \text{STr}\, \pat R_k\,
(\Gamma_k^{(2)}+R_k)^{-1}, \quad t=\ln \frac{k}{\Lambda},
\label{eq:flow_eq1}
\end{equation}
where $\Gamma_k$ interpolates between the bare action
$\Gamma_{k=\Lambda}= S$ and the full quantum effective action
$\Gamma=\Gamma_{k=0}$; $\Gamma_{k}^{(2)}$ denotes the second
functional derivative with respect to the fluctuating field. The
regulator function $R_k$ specifies the Wilsonian momentum-shell
integration, such that the flow of $\Gk$ is dominated by fluctuations
with momenta $p^2\simeq k^2$. 

An approximate solution to the flow equation can reliably describe
also nonperturbative physics if the relevant degrees of freedom in the
form of RG relevant operators are kept in the ansatz for the effective
action. As the crucial ingredient, the choice of this ansatz has to be
guided by all available physical information.

\section{Truncated RG flow for thermal gluodynamics}

In this work, we truncate the space of possible action functionals to
a tractable set of operators which is motivated from various sources
and principles. For the principle of gauge invariance, we use the
background-field formalism as developed in \cite{Reuter:1993kw}, i.e.,
we work in the Landau-de Witt background-field gauge and follow the
strategy of \cite{Reuter:1997gx,Gies:2002af} for an approximate
resolution of the gauge constraints \cite{Ellwanger:iz}. Decomposing
the gauge field into a background-field part and a fluctuation field,
this strategy focusses on the flow in the background-field sector of
the action and neglects an independent running of the
fluctuation-field sector. In general, the solution of the
strongly-coupled gauge sector represents the most delicate part of
this study, owing to a lack of sufficient {\em a priori} control of
nonperturbative truncation schemes. A first and highly nontrivial check
of a solution is already given by the stability of its RG flow, since
oversimplifying truncations which miss the right degrees of freedom
generically exhibit IR instabilities of Landau-pole type.

The IR stability of our solution arises from an important conceptual
ingredient: we optimize our truncated flow with an adjustment of the
regulator to the spectral flow of $\Gamma^{(2)}$
\cite{Gies:2002af,Litim:2002xm}, instead of a naive canonical
momentum-shell regularization. For this, we integrate over shells of
eigenvalues of $\Gamma^{(2)}$, by inserting $\Gamma^{(2)}$ into the
regulator and accounting for the flow of these eigenvalues. More
precisely, we use the exponential regulator \cite{Wetterich:yh} of the
form $R_k(\Gamma^{(2)})=\Gamma^{(2)} /[\exp (\Gamma^{(2)}/\mathcal Z_k
k^2)-1]$, where $\mathcal Z_k$ denotes the wave function
renormalization of the corresponding field (gluons or ghost). In a
perturbative language, the optimizing spectral adjustment allows for a
resummation of a larger class of diagrams.

The main part of our truncation consists of an infinite set of
operators given by powers of the Yang-Mills Lagrangian,
\begin{equation}
\Gamma_{k,\text{YM}}[A]=\int_x \mathcal W_k(\theta), \quad
\theta=\frac{1}{4} F_{\mu\nu}^a F_{\mu\nu}^a. \label{Wtrunc}
\end{equation}
In the function $\mathcal W_k(\theta)=W_1 \theta+ \frac{1}{2} W_{2}
\theta^2+\frac{1}{3!} W_3 \theta^3\dots$, the coefficients $W_i$ form
an infinite set of generalized couplings. This truncation represents a
gauge-covariant gradient expansion in the field strength, neglecting
higher-derivative terms and more complicated color and Lorentz
structures. Hence, the truncation includes arbitrarily high gluonic
correlators projected onto their small-momentum limit and onto the
particular color and Lorentz structure arising from powers of
$F^2$. Since perturbative gluons are certainly not the true degrees of
freedom in the IR, an inclusion of infinitely many gluonic operators
appears mandatory in order to have a chance to capture the relevant
physics in this gluonic language. The covariant gradient expansion
does not only facilitate a systematic classification of the gluonic
operators, it is also a consistent expansion in the framework of the
functional RG.\footnote{The
gluonic gradient expansion as a local expansion can, of course, not be expected to give
reliable answers to all questions; for instance, bound-state phenomena
such as glue balls are encoded in the nonlocal pole structure of
higher-order vertices.} 

Furthermore, our truncation includes standard (bare) gauge-fixing and
ghost terms, neglecting any nontrivial running in these sectors. We
emphasize that we do not expect that this truncation reflects the true
behavior in these sectors, but we assume that the non-trivial running
in these sectors does not qualitatively modify the running of the
background-field sector where we read off the physics. 

A well-known problem of gauge-covariant gradient expansions in
gluodynamics is the appearance of an IR unstable Nielsen-Olesen mode
in the spectrum \cite{Nielsen:1978rm}. At finite temperature $T$, this
problem is severe, since such a mode will be strongly populated by
thermal fluctuations, typically spoiling perturbative computations
\cite{Dittrich:1980nh}. Our flow equation allows us to resolve this
problem with the aid of the IR regulator. We remove this mode's
unphysical thermal population by a $T$-dependent regulator.  In this
way, we obtain a strictly positive spectrum for the thermal
fluctuations.

\begin{figure}[t]
\includegraphics[%
  clip,
  scale=0.7]{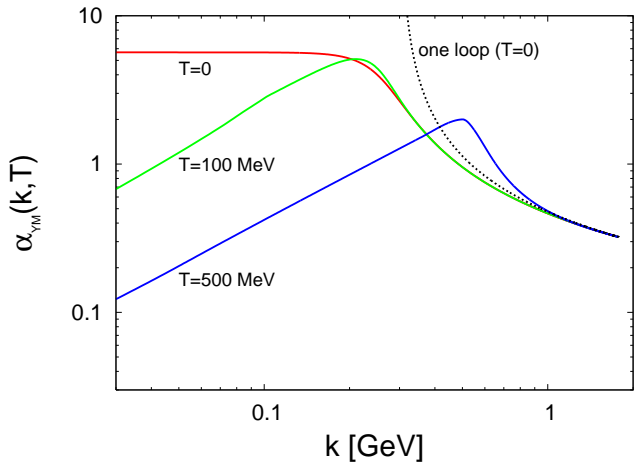}
\caption{\label{coupl_pure_glue} Running $\mathrm{SU}(3)$ Yang-Mills coupling
$\alpha _{\text{YM}}(k,T)$ as a function of $k$ for
$T=0,100,300\,\mathrm{MeV}$ compared to the one-loop 
running for vanishing temperature.}
\label{fig:runcoup}
\end{figure}

In the present truncation, the flow equation results in a differential
equation for the function $\mathcal W_k$ of the form
\begin{equation}
\pat \mathcal W_k(\theta)=\mathcal{F}[\partial_\theta \mathcal W_k,
\partial^2_{\theta} \mathcal W_k,\partial_{t}\partial_{\theta} 
\mathcal W_k,\partial_{t}\partial^2_{\theta}
\mathcal W_k],
\label{I.3}
\end{equation}
where the extensive functional $\mathcal{F}$ depends on derivatives of
$\mathcal W_k$, on the coupling $g$ and the temperature $T$; it is
displayed in \cite{BraunGies}.  We use the nonrenormalization of the
product of coupling and background field, $g A$, for a nonperturbative
definition of the running coupling in terms of the background wave
function renormalization $Z_k\equiv W_1$ \cite{Abbott:1980hw},
\begin{equation}
\beta_{g^{2}}\equiv\partial_{t}g^{2}=\eta\,g^{2}, \quad
\eta=-\frac{1}{Z_k} \pat Z_k. \label{betadef}
\end{equation}
The flow of $Z_k$, and thus the running of the coupling, is successively
driven by all generalized couplings $W_i$. Keeping track of all
contributions from the flows of the $W_i$, Eq.~\eqref{I.3} boils down
to a recursive relation,
\begin{equation}
\pat W_i= f_{ij}(g,T) \pat W_j, \label{eq:rec}
\end{equation}
with ${f}_{ij}(g,T)$ representing the expansion coefficients of the
RHS of Eq.~\eqref{I.3}, which obey $f_{ij}=0$ for $j>i+1$.  Solving
Eq.~\eqref{eq:rec} for $\pat Z_k\equiv \pat W_1$, we
obtain a nonperturbative $\beta_{g^2}$ function in
terms of an infinite asymptotic but resumable power series,
\begin{equation}
\beta_{g^2} = \sum_{m=1}^\infty a_m({\textstyle{\frac{T}{k}}})
\frac{(g^2)^{m+1}}{[2(4\pi)^{2}]^m}, \label{betares}
\end{equation}
with temperature-dependent coefficients $a_m$.\footnote{An iteration
  of the same procedure would result in $\beta$ functions for the
  other higher-order couplings $W_i$ with $i>1$. We neglect these,
  since they do not exert a direct influence on the quark sector
  discussed below.}  For explicit representations of the $a_m$ and
further details, we refer the reader to \cite{BraunGies}.  At zero
$T$, the $\beta_{g^2}$ function agrees well with perturbation theory
for small coupling, reproducing one-loop exactly and two-loop within a
few-percent error.  For larger coupling, the resumed integral
representation of Eq.~\eqref{betares} reveals a second zero of the
$\beta_{g^2}$ function for finite $g^2$, corresponding to an IR
attractive non-Gau\ss ian fixed point $g^2_\ast>0$, which confirms the
results of \cite{Gies:2002af}.  The resulting zero-temperature flow of
the coupling is displayed by the (red) solid ``T=0'' line in
Fig.~\ref{fig:runcoup}. Deviations from perturbation theory become
significant at scales below 1 GeV; the IR fixed-point behavior sets in
on scales of order $\mathcal O$(100 MeV).  As initial condition, we
use the measured value of the coupling at the $\tau$ mass scale
\cite{Bethke:2004uy}, $\alpha_{\mathrm{s}}=0.322$, which evolves to
the world average of $\alpha_{\mathrm{s}}$ at the $Z$ mass scale.  We
stress that no other parameter or scale is used as an input neither at
  $T=0$ nor at finite temperature as described below.

The appearance of an IR fixed point in Yang-Mills theories is a
well-investigated phenomenon in the Landau gauge
\cite{vonSmekal:1997is}, where the running coupling can be defined
with the aid of the ghost-gluon vertex. Whereas the universal
perturbative running of the coupling is identical for the different
gauges despite the different definitions of the coupling, we moreover
observe a qualitative agreement between the Landau gauge and the
Landau-de Witt background-field gauge in the nonperturbative IR in the
form of an attractive fixed point. This points to a deeper connection
between the two gauges which deserves further study and may be traced
back to certain non-renormalization properties in the two gauges. Note
that the IR fixed point in the Landau gauge is in accordance with the
Kugo-Ojima and Gribov-Zwanziger confinement scenarios
\cite{Kugo:1979gm}. In general, an IR fixed point is also compatible
with the existence of a mass gap \cite{Gies:2002af}, since such a gap
in the physical spectrum typically induces threshold and decoupling
behavior towards the IR.

At finite temperature $T$, the UV behavior remains unaffected for
scales $k\gg T$ and agrees well with the one-loop perturbative running
coupling at zero temperature, as expected.\footnote{Our truncation
  does not reproduce higher orders in the high-temperature
  small-coupling expansion which proceeds with odd powers in $g$
  beyond one loop \cite{Kapusta:1979fh}.  These odd powers are a
  result of a resummation which, in the language of the effective
  action, requires non-local operators, being neglected so far. In any
  case, we do not expect the underlying quasi-particle picture of
  these operators to hold near the chiral phase transition, such that
  their omission should not qualitatively modify our low-temperature
  results.} In the IR, the running is strongly modified: The coupling
increases towards lower scales until it develops a maximum near $k\sim
T$. Below, the coupling decreases according to a power law $g^2 \sim
k/T$, see Fig.~\ref{coupl_pure_glue}. This behavior has a simple
explanation: the wavelength of fluctuations with momenta $p^2<T^2$ is
larger than the extent of the compactified Euclidean time direction.
Hence, these modes become effectively 3-dimensional and their limiting
behavior is governed by the spatial 3$d$ Yang-Mills theory. {
  This dimensional reduction has been discussed for the running
  coupling in a perturbative weak-coupling framework in
  \cite{vanEijck:1993zg}. Our results generalize this to arbitrary
  couplings.} As a nontrivial new result, we observe the existence of
a non-Gau\ss ian IR fixed point {$g^2_{3d,\ast}$} also in the
reduced 3-dimensional theory. By virtue of a straightforward matching
between the 4$d$ and 3$d$ coupling, the observed power law for the
4$d$ coupling is a direct consequence of the strong-coupling IR
behavior in the 3$d$ theory, $g^2(k\ll T)\sim g^2_{3d,\ast}\, k/T$.
{We find a 3$d$ fixed-point value of $\alpha_{3d,\ast}\equiv
  g^2_{3d,\ast}/(4\pi)\simeq 2.7$ which demonstrates that the system
  is strongly coupled despite the naive decrease of the 4$d$ coupling;
  also the 3$d$ background anomalous dimension is large, approaching
  $\eta_{3d}\to1$ near the IR fixed point. This scenario is
  reminiscent to hot 4$d$ $\phi^4$ theory which approaches a strongly
  coupled IR limit at high temperatures analogous to the 3$d$
  Wilson-Fisher fixed point \cite{O'Connor:1992cg}.}  The observation
of an IR fixed point in the 3$d$ theory again agrees with recent
results in the Landau gauge \cite{Maas:2005hs}. {Note that our
  flow to the 3$d$ theory is driven by thermal as well as quantum
  fluctuations. This is different from a purely thermal flow
  \cite{D'Attanasio:1996zt,Litim:1998nf} as
  used in \cite{Comelli:1997ru}, where the IR limit is
  four-dimensional, being characterized by a decoupling of fluctuation
  modes owing to thermal masses. }

The 3$d$ IR fixed point and the perturbative UV behavior already
qualitatively determine the momentum asymptotics of the running
coupling. Phenomenologically, the behavior of the
coupling in the transition region at mid-momenta is most important,
which is quantitatively provided by the full 4$d$
finite-temperature flow equation.

\section{Truncated RG flow for chiral quark dynamics}

Extending our calculations to QCD, we first include the quark
contributions to all gluonic operators of our truncation, as done in Ref. 
\cite{Gies:2004hy} for QED. This effectively corresponds to
Heisenberg-Euler-type quark-loop contributions to the flow of the
function $\mathcal W_k(\theta)$.  Successively, we obtain quark-loop
contributions to the coefficients $a_m$ in Eq.~\eqref{betares} and
thus to the running coupling, accounting for the screening nature of
fermionic fluctuations; here, we confine ourselves to massless quarks,
but current-quark masses can straightforwardly be included~\cite{BraunGies}.  

The determination of the critical temperature
$\Tc$ above which chiral symmetry is restored requires a second
crucial ingredient for our truncation: we study the
gluon-induced quark self-interactions of the type
\begin{equation}
\Gamma_{\psi,\text{int}}=\int \hat\lambda_{\alpha\beta\gamma\delta}
\yb_\alpha \psi_\beta \yb_\gamma \psi_\delta, \label{psitrunc}
\end{equation}
where $\alpha,\beta,\dots$ denote collective indices including color,
flavor, and Dirac structures.  The resulting flow equations for the
$\hat\lambda$'s are a straightforward finite-temperature
generalization of those derived and analyzed in
\cite{Gies:2003dp,Gies:2005as} and will not be displayed here for
brevity. The boundary condition
$\hat\lambda_{\alpha\beta\gamma\delta}\to 0$ for
$k\to\Lambda\to\infty$ guarantees that the $\hat\lambda$'s at
$k<\Lambda$ are solely generated by quark-gluon dynamics from first
principles (e.g., by 1PI ``box'' diagrams with 2-gluon exchange). 
We emphasize that this is an important difference to, e.g., the 
Nambu-Jona-Lasinio (NJL) model, where the $\hat\lambda$'s are 
independent input parameters.

We consider all linearly-independent four-quark interactions permitted
by gauge and chiral symmetry. A priori, these include color and flavor
singlets and octets in the (S$-$P), (V$-$A) and (V$+$A) channels.
U${}_{\text{A}}(1)$-violating interactions are neglected, since they
may become relevant only inside the \xsb\ regime or for small $\Nf$.
We drop any nontrivial momentum dependencies of the $\hat\lambda$'s
and study these couplings in the point-like limit
$\hat\lambda(|p_i|\ll k)$. This is a severe approximation, since it
inhibits a study of QCD properties in the chirally broken regime; for
instance, mesons manifest themselves as momentum singularities in the
$\hat\lambda$'s.  Nevertheless, the point-like truncation can be a
reasonable approximation in the chirally symmetric regime, as has
recently been quantitatively confirmed for the zero-temperature chiral
phase transition in many-flavor QCD \cite{Gies:2005as}. Our truncation
is based on the assumption that quark dynamics both near the
finite-$T$ phase boundary as well as near the many-flavor phase
boundary \cite{Banks:1981nn} is driven by similar mechanisms. Our 
restrictions on the four-quark interactions result in a total number of 
four linearly-independent $\hat\lambda$ couplings; all others channels are 
related to this minimal basis by means of Fierz transformations.

Introducing the dimensionless couplings $\lambda=k^2 \hat\lambda$, the
$\beta$ functions for the $\lambda$ couplings are of the form
\begin{equation}
\pat\lambda=2\lambda - \lambda A \lambda - b \lambda g^2 - c g^4, 
\label{eqlambda}
\end{equation}
where the coefficients $A$, $b$, $c$ are temperature dependent, $A$ being
a matrix and $b$ a vector in the space of $\lambda$ couplings {(for 
explicit representations, see \cite{BraunGies,BraunDiss})}. 

Within this truncation, a simple picture for the chiral dynamics
arises: at weak gauge coupling, the RG flow generates quark
self-interactions of order $\lambda\sim g^4$ via the last term in
Eq.~\eqref{eqlambda} with a negligible back-reaction on the gluonic RG
flow. If the gauge coupling in the IR remains smaller than a critical
value $g<g_{\text{cr}}$, the $\lambda$ self-interactions remain
bounded, approaching fixed points $\lambda_\ast$ in the IR.
Technically, the $\sim g^4$ term is balanced by the first term
$\sim 2\lambda$ at these fixed points.  The fixed points are the
counter-parts of the Gau\ss ian fixed point
$\lambda_\ast^{\text{Gau\ss}}=0$ in NJL-like models (at $g^2=0$), here
being modified by the gauge dynamics. At these fixed points, the
fermionic subsystem remains in the chirally invariant phase which is
indeed realized at high temperatures $T>T_{cr}$.

If the gauge coupling increases beyond the critical coupling
$g>g_{\text{cr}}$, the IR fixed points $\lambda_\ast$ are destabilized
and the quark self-interactions become critical
\cite{Gies:2003dp,Gies:2005as}. Then the gauge-fluctuation-induced
$\lambda$'s have become strong enough to contribute as relevant
operators to the RG flow, with the term $\sim \lambda A\lambda$
dominating Eq.~\eqref{eqlambda}. In this case, the $\lambda$'s
increase rapidly, approaching a divergence at a finite scale
$k=k_{\text{\xsb}}$. In fact, this seeming Landau-pole behavior
indicates \xsb\ and the formation of chiral condensates: the
$\lambda$'s are proportional to the inverse mass parameter of a
Ginzburg-Landau effective potential for the order parameter in a
(partially) bosonized formulation, $\lambda\sim 1/m^2$
\cite{Ellwanger:1994wy,Gies:2001nw}. Thus, the scale at which the
self-interactions formally diverge is a good measure for the scale
$k_{\text{\xsb}}$ where the effective potential for the chiral order
parameter becomes flat and is about to develop a nonzero vacuum
expectation value.

Whether or not chiral symmetry is preserved by the ground state
therefore depends on the running QCD coupling $g$ relative to the critical
coupling $g_{\text{cr}}$ which is required to trigger \xsb. For
instance, at zero temperature, the SU(3) critical coupling for the
quark system is $\alpha_{\text{cr}}\equiv g^2_{\text{cr}}/(4\pi)\simeq
0.8$ in our RG scheme \cite{Gies:2001nw}, being only weakly dependent
on the number of flavors \cite{Gies:2005as}.  Since the IR fixed point
for the gauge coupling is much larger $\alpha_\ast>\alpha_{\text{cr}}$
for not too many massless flavors, the QCD vacuum is characterized by
\xsb.
\begin{figure}[t]
\includegraphics[%
  clip,
  scale=0.7]{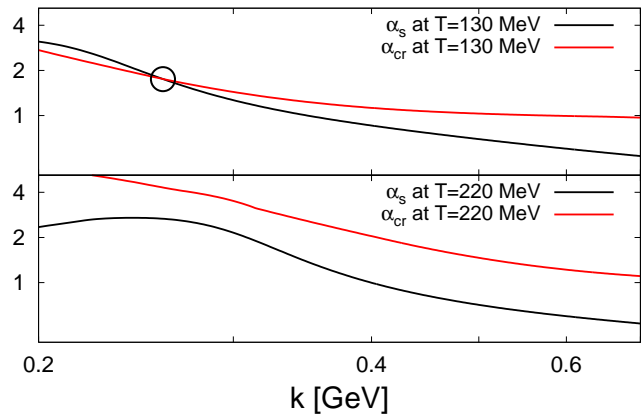}
\caption{\label{alpha_alpha_c} Running QCD coupling
$\alpha_{\mathrm{s}} (k,T)$ for $\Nf=3$ massless quark flavors and
$\Nc=3$ colors and the critical value of the running coupling
$\alpha_{\mathrm{cr}} (k,T)$ as a function of $k$ for
$T=130\,\mathrm{MeV}$ (upper panel) and $T=220\,\mathrm{MeV}$ (lower
panel). The existence of the
$(\alpha_{\mathrm{s}},\alpha_{\mathrm{cr}})$ intersection point (marked
by a circle) in the former indicates that the \xsb\ quark dynamics can 
become critical.}
\end{figure}
At finite temperature, the running of the gauge coupling is
considerably modified in the IR. Moreover, the critical coupling is
$T$ dependent, $g_{\text{cr}}=g_{\text{cr}}(T/k)$. This can be
understood from the fact that all quark modes acquire thermal masses
and, thus, stronger interactions are required to excite critical quark
dynamics. This thermal decoupling is visible in the coefficients $A$,
$b$, and $c$ in Eq.~\eqref{eqlambda}, all of which vanish in the limit
$T/k\to \infty$. 

In Fig. \ref{alpha_alpha_c}, we show the running coupling
$\alpha_{\mathrm{s}}$ and its critical value $\alpha_{\mathrm{cr}}$
for $T=130\,\mathrm{MeV}$ and $T=220\,\mathrm{MeV}$ as a function of
the regulator scale $k$. The intersection point $k_{\text{cr}}$
between both marks the scale where the quark dynamics become
critical. Below the scale $k_{\text{cr}}$, the system runs quickly
into the \xsb\ regime. We estimate the critical temperature
$T_{\text{cr}}$ as the lowest temperature for which no intersection
point between $\alpha_{\mathrm{s}}$ and $\alpha_{\mathrm{cr}}$
occurs.\footnote{Strictly speaking, this is a sufficient but not a
necessary criterion for chiral-symmetry restoration. In this sense,
our estimate for $T_{\text{cr}}$ is an upper bound for the true
$T_{\text{cr}}$. Small corrections to this estimate could arise, if
the quark dynamics becomes uncritical again by a strong decrease of
the gauge coupling towards the IR.}  { Compared to
\cite{BraunGies}, we have further resolved the finite-$T$ Lorentz
structure of the four-fermion couplings \cite{BraunDiss}, resulting in
a slightly improved estimate for $T_{\text{cr}}$: we find
$T_{\mathrm{cr}}\approx 172\genfrac{}{}{0pt}{}{+ 40}{-
34}\,\mathrm{MeV}$ for $\Nf=2$ and $T_{\mathrm{cr}}\approx
148\genfrac{}{}{0pt}{}{+ 32}{- 31}\,\mathrm{MeV}$ for $\Nf=3$ massless
quark flavors in good agreement with lattice
simulations~\cite{Karsch:2000kv}.  The errors arise from the
experimental uncertainties on $\alpha_{\mathrm{s}}$
\cite{Bethke:2004uy}.  Dimensionless ratios of observables are less
contaminated by this uncertainty of $\alpha_{\text{s}}$. For instance,
the relative difference for $\Tc$ for $\Nf$=2 and 3 flavors is
$\frac{\Tc^{\Nf=2}-\Tc^{\Nf=3}}{(\Tc^{\Nf=2}+\Tc^{\Nf=3})/2}=0.150\dots
0.165$ in reasonable agreement with the lattice
value\footnote{{The large uncertainty on the lattice value arises
from the fact that the statistical errors on the $\Nf=2$ and $\Nf=3$
results for $T_{\text{cr}}$ are uncorrelated.}}  of $\sim 0.121\pm
0.069$.}

\begin{figure}[t]
\includegraphics[%
  clip,
  scale=0.7]{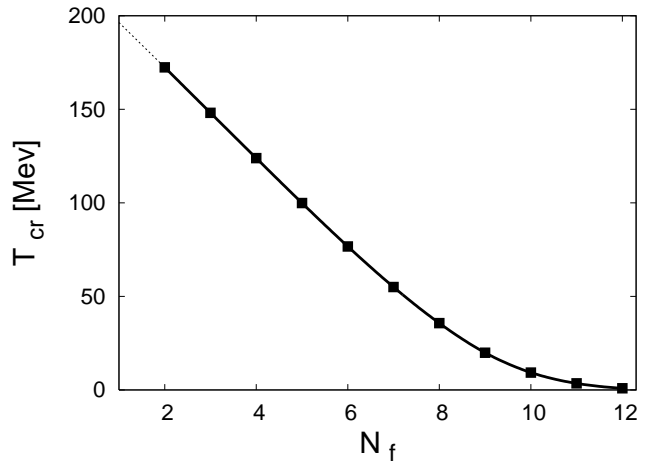}
\caption{\label{tc_nf} Chiral-phase-transition temperature
$T_{\text{cr}}$ versus the number of massless quark flavors $\Nf$. In
the dashed-line region, we expect U${}_{\text{A}}(1)$-violating
operators to become quantitatively important. The flattening at
$\Nf\gtrsim10$ is a consequence of the IR fixed-point {structure~\cite{BraunGies}.}}
\end{figure}

Furthermore, we compute the critical temperature for the case of many
massless quark flavors $\Nf$, see Fig.~\ref{tc_nf}. We observe an
almost linear decrease of the critical temperature for increasing
$\Nf$ with a slope of {$\Delta T_{\mathrm{cr}}=T(\Nf)-T(\Nf+1)\approx
24\,\mathrm{MeV}$} for small $\Nf$.  In addition, we find a critical
number of quark flavors, $\Nf ^{\mathrm{cr}}=12$, above which no
chiral phase transition occurs.  This result for $\Nf^{\mathrm{cr}}$
agrees with other studies based on the 2-loop $\beta$ function
\cite{Banks:1981nn}; however, the precise value of $\Nf^{\mathrm{cr}}$
is {exceptionally} sensitive to the 3-loop coefficient which can
bring $\Nf^{\text{cr}}$ down to $\Nf^{\text{cr}}\simeq
10{\genfrac{}{}{0pt}{}{+ 1.6}{- 0.7}}$ \cite{Gies:2005as}.
{Since we do not consider our truncation to be sufficiently
  accurate for a precise estimate of this coefficient, our study does
  not contribute to a reduction of the current error on
  $\Nf^{\text{cr}}$. Instead, w}e would like to emphasize that the
flattening shape of the phase boundary near $\Nf^{\text{cr}}$ is a
generic prediction of the IR fixed-point scenario: here, the symmetry
status of the system is governed by the fixed-point regime where
dimensionful scales such as $\Lambda_{\text{QCD}}$ lose their
importance \cite{BraunGies}. In any case, since $\Nf ^{\mathrm{cr}}$
is smaller than $\Nf ^{\mathrm{a.f.}}=\frac{11}{2} N_{c}=16.5$, our
study {provides further evidence for} the existence of a regime
where QCD is chiral symmetric but is still asymptotically free.

\section{Conclusion}
 
In summary, we have determined the \xsb\ phase boundary in QCD in the
plane of temperature and flavor number. Our quantitative results are
in accord with lattice simulations for $\Nf=2,\, 3$. For larger $\Nf$, we
observe a linear decrease of $\Tc$, leveling off near
$\Nf^{\text{cr}}$ owing to the IR fixed-point structure of QCD. Our
results are based on a consistent operator expansion of the QCD
effective action that can systematically be generalized to higher
orders. 

The qualitative validity and the quantitative convergence of this
expansion are naturally difficult to analyze in this strongly-coupled
gauge system, particularly for the gluonic sector. The fact that our
truncation results in a stable RG flow at strong interactions is
already a highly non-trivial check that any ansatz which misses the
true degrees of freedom generically fails. A more quantitative
evaluation of the validity of our expansion will require the inclusion
of higher-order operators in the covariant gradient expansion as well
as higher-order ghost terms. {An inclusion of operators that
distinguish between electric and magnetic sectors at finite $T$,
e.g., {$(u_\mu F_{\mu\nu})^2$} with the heat-bath four-velocity $u_\mu$,
should facilitate to distinguish between differing coupling strengths
in the two sectors, as done in \cite{Comelli:1997ru} using an
{ansatz inspired by hard thermal loop computations.}} 

We observe an improved control over the truncation in the quark sector
at least for the chirally symmetric phase, which suffices to trace out
the phase boundary. Quantitatively, this has been confirmed by a
stability analysis of universal quantities such as $\Nf^{\text{cr}}$
under a variation of the regulator in \cite{Gies:2005as} which gives
strong support to the point-like truncation of the quark
self-interactions. Qualitatively, the reliability of the quark
truncation can also be understood by the fact that the feed-back of
higher-order operators, such as $\sim(\yb\psi)^4$, is generally
suppressed by the one-loop structure of the flow equation.

Future extensions should include mesonic operators which can
be treated by RG rebosonization techniques \cite{Gies:2001nw}.  This
would not only provide access to the broken phase and mesonic
properties, but also permit a study of the order of the phase
transition. For further phenomenology, the present quantitative
results that rely on only one physical input parameter can serve as a
promising starting point.

The authors are grateful to J.~Jaeckel, J.M. Pawlowski, and
H.-J.~Pirner for useful discussions.  H.G.~ acknowledges support by
the DFG under contract Gi 328/1-3 (Emmy-Noether
program). J.B. acknowledges support by the GSI Darmstadt.

\end{document}